\documentclass[usenatbib]{mn2e}
\usepackage[dvips]{graphicx}

\title[Comparison of the atmosphere in Antarctica]{Comparison of the
  atmosphere above the South Pole, Dome C and Dome A: first attempt}
\author[S. Hagelin et al.]{S. Hagelin,$^{1, 2}$\thanks{E-mail:
     hagelin@arcetri.astro.it; masciadri@arcetri.astro.it} E. Masciadri$^1$\footnotemark[1],
F. Lascaux$^1$ and J. Stoesz$^1$ \\ $^1$INAF Osservatorio Astrofisico
di Arcetri, Largo Enrico Fermi 5, I-501 25 Florence, Italy\\
$^2$Uppsala Universitet, Department of Earth Sciences, Villav\"agen 16,
S-752 36 Uppsala, Sweden}

\begin{document}
\label{firstpage}
\date{Accepted 2008 April 17, Received 2007 April 3; in original form
2007 November 14}  
\pagerange{\pageref{firstpage}--\pageref{lastpage}}
\pubyear{2007}

\maketitle

\begin{abstract}
The atmospheric properties above three sites (Dome C, Dome A and the
South Pole) on the Internal Antarctic Plateau are investigated for
astronomical applications using the monthly median of the analyses
from ECMWF (the European Centre for Medium-Range Weather
Forecasts). Radiosoundings extended on a yearly time scale at the South
Pole and Dome C are used to quantify the reliability of the ECMWF
analyses in the free atmosphere as well as in the boundary and surface
layers, and to characterize the median wind speed in the first 100 m
above the two sites. Thermodynamic instability properties in the free
atmosphere above the three sites are quantified with monthly median
values of the Richardson number. We find that the probability to
trigger thermodynamic instabilities above 100 m is smaller on the
Internal Antarctic Plateau than on mid-latitude sites. In spite of
the generally more stable atmospheric conditions of the Antarctic
sites compared to mid-latitude sites, Dome C shows worse thermodynamic
instability conditions than those predicted above the South Pole and
Dome A above 100 m. A rank of the Antarctic sites done with
respect to the strength of the wind speed in the free atmosphere
(ECMWF analyses) as well as the wind shear in the surface layer
(radiosoundings) is presented.
\end{abstract}


\section{Introduction}
The summits of the Antarctic Plateau are of particular interest for
ground-based astronomy because the optical turbulence appears to be
confined to a narrow layer close to the ground,
\citep{ma2,la,ag}. Above this layer the atmosphere is exceptionally
clear and the turbulence weak (0.27\arcsec \cite{la}, 0.36\arcsec
\cite{ag} at Dome C). Measurements have shown that the height of the
turbulent layer above the summits is much lower than above other sites
on the Plateau, \citep{la, maa, ma, ma2}. More precisely, the height of
this layer is much larger above the South Pole (220 m \citep{ma} or
270 m \citep{tr}), which lies on a slope, than above Dome C (36
$\pm$10 m \cite{ag}). Above Dome A the turbulence is expected to be
even weaker.

The surface winds of Antarctica are among the principal sources of
turbulence near the ground. The dominant source of the surface winds
is the sloping terrain and the radiative cooling of the surface
\citep{sch}. The radiative cooling produces a temperature inversion
that together with the sloping terrain cause a horizontal temperature
gradient. This triggers a surface wind alongside the slope of the
terrain. Strong wind shears can therefore occur in the boundary
between the surface winds and the winds in the free atmosphere, which
in general are geostrophic. This is the main source of instability
under conditions of extreme stability, as is the case of the Antarctic
atmosphere in winter. Above the summits of the Internal Antarctic
Plateau the surface winds should be much weaker than elsewhere on the
Plateau due to a lack of the principal cause triggering them: a
sloping terrain.

These elements justify the enthusiastic interest of astronomers for
this site \citep{sto,fos}
Studies of the atmospheric 
properties of these regions are
fundamental for applications to ground-based astronomy. However we need 
a better quantification of these characteristics, using instruments and
measurements as well as models and simulations, in order to fill the
gaps of uncertainties or doubts that still remain \citep{forot}
, extending our
attention to a comparative analysis of different sites of the Internal
Antarctic Plateau.

It is important to produce statistical estimate of meteorological
parameters at all heights from the ground to verify if atmospheric
conditions are {\it always} advantageous for astronomical
applications.  Indeed, it has recently been shown \citep{gm}, using
European Centre for Medium-Range Weather Forecasts (ECMWF) analyses,
that in winter the wind speed grows monotonically above $\sim$10 km
a.s.l. (above sea level), achieving median values of the order of
$\sim$30 m s$^{-1}$ at 25 km a.s.l.  At this height a variation of
$~\sim$20 m s$^{-1}$ has been estimated between summer and
winter. Such a strong wind might trigger an important decrease of the
wavefront coherence time, even in presence of a weak turbulence (see
discussion in \cite{gm}) and, as a consequence, the potentiality of
these sites might vanish. It should be therefore interesting to better
quantify the median wind speed profile on other sites (with
astronomical interest) of the Internal Antarctic Plateau or to
retrieve some general indication of the wind speed in the free
atmosphere above the Internal Antarctic Plateau.

Besides that, the employment of ECMWF analyses for characterization of
the surface layer requires a deeper analysis. \cite{gm} 
 concentrated their study at heights greater
than 30 m thus excluding the surface contribution, assuming that the
ECMWF analyses are not optimized for the atmospheric flow near the
surface. More recently, studies \citep{sa} 
appeared claiming that the ECMWF analyses can be used to quantify and
characterize the atmospheric properties with a good level of accuracy
down to the surface level.  In spite of our conviction that this
conclusion is the result of a partial analysis (only summer data) we
admit that in \cite{gm} 
the authors {\it assumed} (and they did not proved) the limits of the
ECMWF for the surface layer.  We therefore think that is time to
provide a dedicated analysis on this subject to know the limit within 
which we can achieve reliable estimates with General Circulation
Models (GCM) i.e. with ECMWF analyses and to identify the domains in 
which one is forced to employ mesoscale models. The latter are 
in principle more suitable to better resolve phenomena happening at 
smaller spatial and temporal scales.

 The usefulness of mesoscale models depends on the limitations imposed
by the products of the General Circulation Models (that means the
ECMWF analyses). It is obvious that, the usefulness of mesoscale
models would not be justified for wind speed if ECWMF products could
provide answers with a sufficient good accuracy.

Our group is involved in a long term study made with meso-scale models
for the simulation of the optical turbulence on the whole troposphere
and low stratosphere using the technique described in \cite{b17}
above the Internal Antarctic
Plateau\footnote{Some attempts have been done in the past \citep{sg}
even if with different scientific goals.}. This study is
therefore propedeutic to researches done with such a typology of
models.

It is therefore fundamental to provide a clear picture of the
limitations of the ECMWF products and at the same time, to try to
retrieve the maximum number of useful information we can get from such
a kind of products.

In this paper we try a first attempt to quantify, above the three
sites with some astronomic interests (the South Pole, Dome C and Dome
A), the differences of some critical meteorologic parameters that are
directly, or indirectly, related to the characteristics of atmospheric
turbulence. We expressly select two sites (South Pole and Dome C) for
which measurements are available and one site (Dome A) for which no
measurements are available.  The reasons for this choice is explained
later on (paper's scientific goals).

We use data from the MARS catalogue of the ECMWF and radiosoundings
from the South
Pole\footnote{ftp://amrc.ssec.wisc.edu/pub/southpole/radiosonde} and
Dome C\footnote{http://www.climantartide.it}. Analyses data are
extracted from the three grid points that are nearest to the sites of
interest, i.e. Dome A, Dome C and the South Pole. The coordinates of
the sites are given in Table \ref{ou}. We extracted analyses data from
MARS at 00:00 UTC for the whole year of 2005 to assure a complete
statistical sample covering all seasons. A more detailed description
of the analyses data set is given by \cite{gm}.\newline

\begin{table}
\caption{The geographic coordinates of the sites and the closest grid
points from which the ECMWF analyses are extracted.}
\begin{tabular}{llr@{\degr}l} 
\hline
Site & Lat. & \multicolumn{2}{l}{Long.} \\
\hline
Dome A$^*$      & 80\degr22\arcmin00\arcsec S &  77 & 21\arcmin11\arcsec E \\
                & 80\degr30\arcmin00\arcsec S &  77 & 30\arcmin00\arcsec E \\
Dome C $^{**}$  & 75\degr06\arcmin04\arcsec S & 123 & 20\arcmin48\arcsec E \\
                & 75\degr00\arcmin00\arcsec S & 123 & 30\arcmin00\arcsec E \\
South Pole      & 90\degr00\arcmin00\arcsec S &   0 & 00\arcmin00\arcsec E \\
                & 90\degr00\arcmin00\arcsec S &   0 & 00\arcmin00\arcsec E \\
\hline
\end{tabular}
\medskip
{\\$^*$Measured with GPS by Dr X. Cui (private communication).
\\$^{**}$Measured with GPS by Prof. J. Storey (private communication).}
\label{ou}
\end{table}

The scientific goals of this paper are:
\begin{enumerate}
\renewcommand{\theenumi}{(\arabic{enumi})}
\item To carry out a detailed comparison of radiosoundings/analyses of
  the wind speed and the potential temperature (the main critical
  parameters defining the stability of the atmosphere) near the
  surface (the first 150 m) for winter as well as summer above Dome C
  and the South Pole. This will permit us to quantify the uncertainty
  between measurements and ECMWF analyses in this vertical slab.  The
  idea is therefore to define the conditions in which the ECMWF can be
  used to characterize, with a good level of reliability and accuracy,
  some atmospheric parameters and to use this tool to characterize a
  site for which no measurements are available.  This is of course the
  interest for a model. Depending on the results of this analysis we
  will perform comparisons of meteorologic parameters above the South
  Pole, Dome C and Dome A using ECMWF analyses (Section 2).
\item Using radiosoundings we will estimate the statistic median
  values of the wind speed in the first tens of meters above the South
  Pole and Dome C extended from April to November. We will therefore be
  able to quantify which site shows the better characteristics for
  nightly astronomical applications (Section 3).
\item We extend the study developed by \cite{gm} above Dome C for the
  ECMWF wind speed also to the South Pole and Dome A, both located on
  the Internal Antarctic Plateau but at different latitude and
  longitude. In this way, we intend to quantify which site is the
  best for astronomical applications. Results of this analysis are fundamental 
  to confirm or see in the right perspective the potentialities of Dome C.
\item We extend the analysis of the Richardson number done by
  \cite{gm} for Dome C for heights above 30 m to the three sites
  (South Pole, Dome C and Dome A) in order to quantify the regions and
  the periods that are less favorable for the triggering of optical
  turbulence and to identify the site with the best characteristics
  (Section 5). This result should represent the first estimate of
  potentialities of Dome A for astronomical applications and this
  should mean that we are able to provide some reliable results and
  conclusions even before some measurements are done on that site.
  This study has therefore a double interest. Firstly the intrinsic
  result itself. Secondly this analysis should open the path to a
  different approach for a fast and reliable classifications of
  potential astronomical sites.
\end{enumerate}

\section{ECMWF analyses versus radiosoundings}

\begin{figure*}
\includegraphics[]{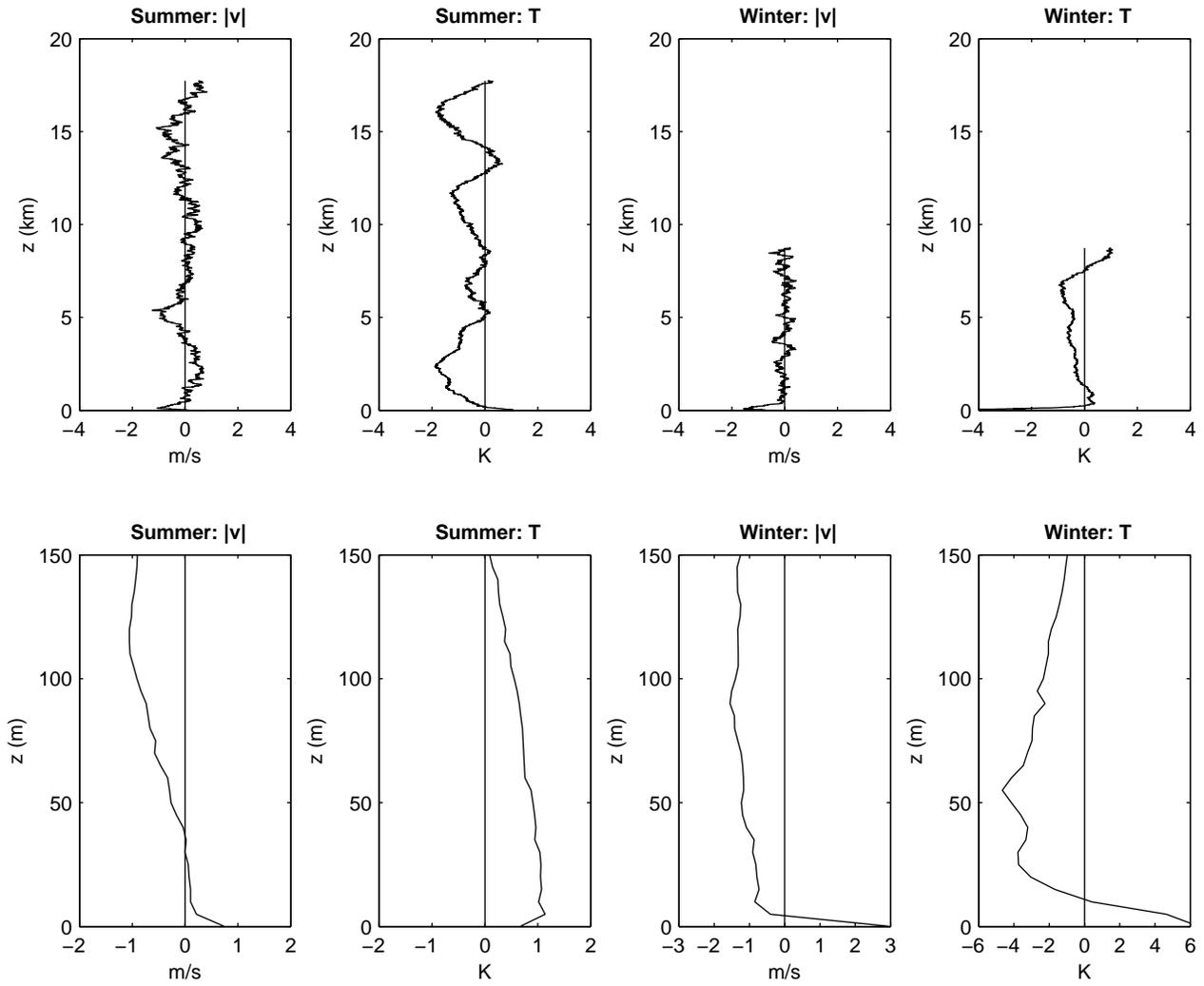}
\caption{The median of the difference of the absolute temperature and
  the wind speed between ECMWF analyses and radiosoundings (ECMWF - radiosoundings) for summer
  (December, January, February) and winter (June, July, August) at
   Dome C in 2006. 
\label{motd}}
\end{figure*}

\begin{figure*}
\includegraphics[]{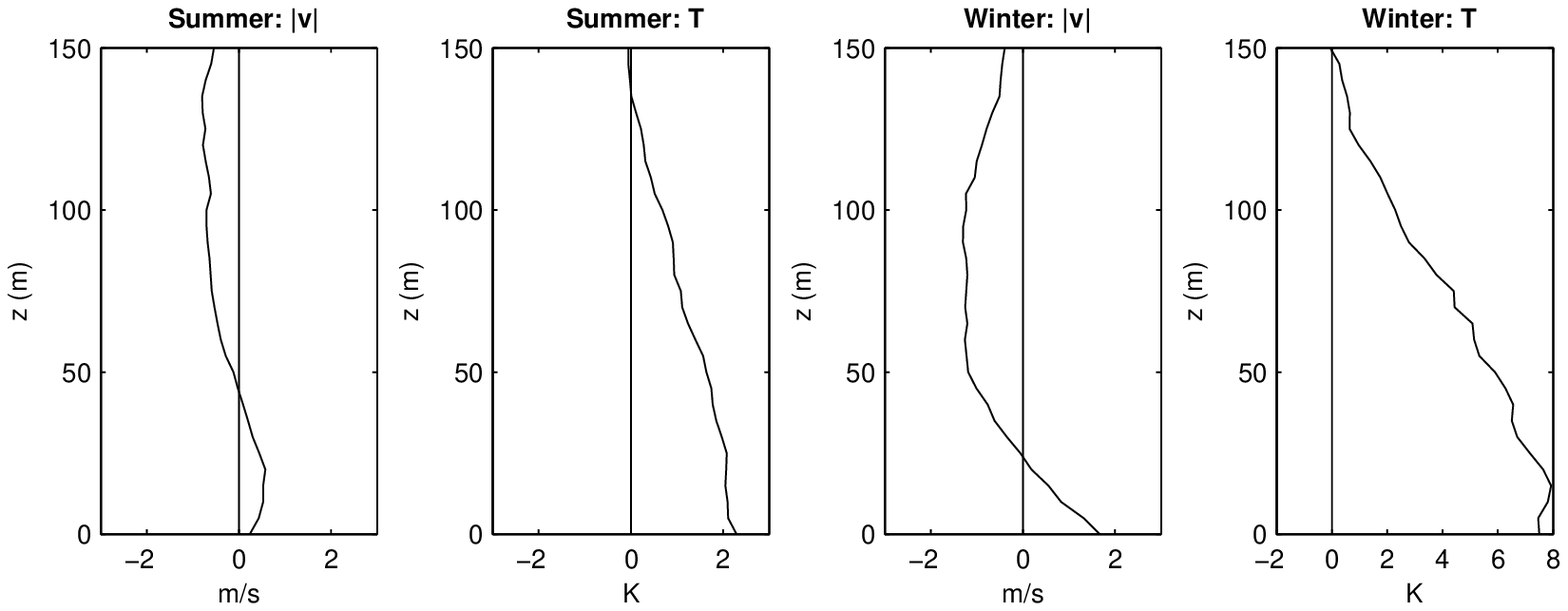}
\caption{The median of the difference of the absolute temperature and
  the wind speed between ECMWF analyses and radiosoundings (ECMWF - radiosoundings) for summer
  (December, January, February) and winter (June, July, August) at
  South Pole in 2006.
\label{motdd}}
\end{figure*}

\begin{figure*}
\vbox to120mm{ 
\includegraphics{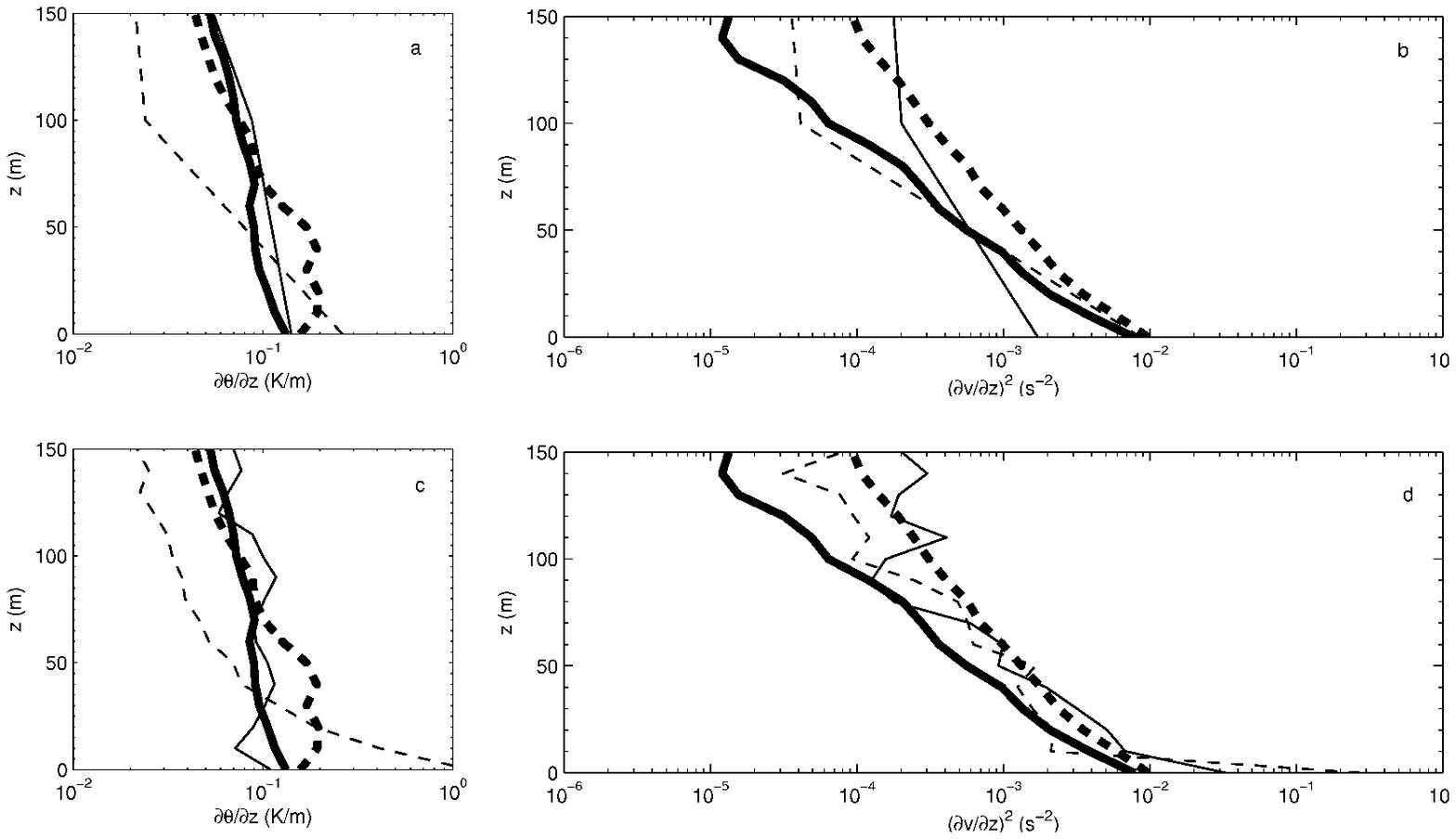}
\caption{The gradients of the potential temperature and the wind speed
  near the surface for Dome C and the South Pole in July 2006. The dashed lines
  refer to Dome C and the solid lines refer to the South Pole. Thick
  lines are ECMWF analyses and thin lines are radiosoundings. In the
  top plots (a and b) the radiosoundings are interpolated with a step
  of 100 m, in the bottom plots (c and d) the step is 10 m.\label{rs1}}
}
\end{figure*}

In this section the ECMWF analyses are compared to radiosoundings from
Dome C and the South Pole in order to investigate the reliability of
the ECMWF data over an antarctic site.  The radiosoundings and
the ECMWF analyses used for this comparison refer to the year 2006. This
year was chosen because of the richer sample of available
radiosoundings. The reason why the comparison of analyses from
different sites, discussed later on in this paper, are from 2005 is
because we wish to investigate a homogeneous data set. In February
2006 the number of vertical levels in the ECMWF model changed from 60
to 91.

When studying the difference between radiosoundings and analyses
particular interest is paid to the first tens of meters since this is
where we expect the largest turbulent activity and this range was not
studied by \cite{gm}. The median of the difference in wind speed and
temperature between ECMWF analyses and radiosoundings, for summer
(December, January and February) and winter (June, July and August)
are shown in Fig. \ref{motd}.  A total number of 73 nights has been
used in summer and 75 in winter. In the free atmosphere we observe
that the radiosoundings do not reach as high altitudes in winter as
they do in summer.  A similar effect has been observed above the South
Pole in our previous study \citep{gm} 
This effect makes it difficult to study the reliability of the ECMWF
data at high altitudes in winter. In this season the balloons
frequently explode. This happens, highly probably, because, due to
the low pressure in the high part of the atmosphere, the balloons
increase their size and they finally explode. The elastic material of
the balloons is much more fragile in winter, when the temperature is
much lower than in summer in the high part of the atmosphere.

In local summer the median difference in the wind speed never exceeds
1 m s$^{-1}$. Closest to the ground the difference is even smaller,
below 1 m s$^{-1}$. During local winter the median difference in wind speed
 never exceeds 0.5 m s$^{-1}$ in the upper atmosphere, though
the radiosoundings only reach $\sim$10 km above the ground. The median
difference is larger closest to the ground ($\sim$ 3 m
s$^{-1}$) in winter.

The median absolute temperature difference in summer is below 2 K throughout
the whole 20 km. In the proximity of the surface this difference is of
the order of 1 K, closest to the surface even less. During the winter
the median difference is similar to that of the summer in the high
part of the atmosphere. In the first 100 m the median difference is
significantly larger, more than 6 K nearest the surface.

\begin{table*}
\caption{The median wind speed and absolute temperature for the winter months
  (JJA) from the lowest level of the radiosoundings compared to values
  obtained with AWS. First and third quartile are also given.}
\begin{tabular}{lcccccc}
\hline
 &  \multicolumn{3}{c}{v$_0$ (ms$^{-1})$} & \multicolumn{3}{c}{T$_0$
 (K)} \\
  & 1$^{st}$ quartile & median & 3$^{rd}$ quartile &  1$^{st}$ quartile &
  median & 3$^{rd}$ quartile \\
\hline
 & \multicolumn{3}{c}{Radiosoundings} & \multicolumn{3}{c}{Radiosoundings}   \\
Dome C (2006)     & 1.5 & 2.8 & 3.8 & 204.4 & 208.3 & 213.1 \\
South Pole (2004) & 4.6 & 5.7 & 7.2 & 205.8 & 209.7 & 213.6 \\
\\
  & \multicolumn{3}{c}{AWS} & \multicolumn{3}{c}{AWS}     \\
Dome C (2006)     & 2.0 & 3.0 & 4.5 & 204.8 & 209.0 & 213.6 \\
South Pole (2004)  & 3.6 & 5.7 & 7.4 & 206.7 & 211.4 & 215.0 \\
\hline
\end{tabular}

\label{aws}
\end{table*}

Figure \ref{motdd} shows the same outputs of Fig.\ref{motd} but
calculated for South Pole. We report just the first 150 m because we
had already calculated \citep{gm} 
the difference ECMWF analyses/radiosoundings above 150 m at South Pole.
For the wind speed the conclusion is similar to what observed at Dome
C. The wind speed difference remains within 1 m s$^{-1}$ in the first
150 m in summer. However, analyses show a tendency in overestimating
($\sim$ 2 m s$^{-1}$) the wind speed in the very low levels in winter
even if this effect is slightly weaker than above Dome C. The median
absolute temperature difference in summer is similar to what observed
above Dome C and contained within $\sim$ 1 K. Near the ground the
ECMWF analyses are almost 2 K warmer than the radiosoundings. This
same trend is observed also in winter. However, in this season, the
analyses are visibly much warmer ($\sim$ 6 K) than the radiosoundings
near the surface.  The statistic uncertainty $\sigma/\sqrt N$ for the
wind speed is of the order of $0.2$ m s$^{-1}$ all along the 20 km
with a maximum of $0.4$ m s$^{-1}$ at 5 km in summer. The
statistic uncertainty for the absolute temperature is of the order of
$0.3$ K and slightly larger ($\sim$ $0.6$ K) in the first 30 m in
winter. The precision of the tempearture provided by radiosoundings is
$\sim$ $0.2$ K in the boundary layer and $\sim$ $0.4$ K in the free
atmosphere while the precision of the wind speed\footnote{More
precisley, the velocity uncertainty is defined in the technical
specification as the standard deviation of the differences in twin
soundings.} is of the order of $0.15$ m s$^{-1}$.

Summarizing, we should say that the wind speed is almost well
reconstructed by the ECMWF analyses with exception of the surface in
winter where ECMWF analyses show a tendency in overestimating of 2-3 m
s$^{-1}$. For a typical wind speed of 3 m s$^{-1}$ this corresponds to
a large discrepancy. The absolute temperature, is in general warmer
from the ECMWF analyses than from the radiosoundings near the surface
in winter achieving a difference of the order of $\sim$ 6 K. The wind
speed and temperature show similar trends above the two sites with
exception of the absolute temperature in winter. At South Pole, the
temperature from ECMWF analyses appears warmer on the whole 150 m
while, at Dome C, only in the first 20 m.

To check if any biases are present in the measurement provided by
radiosoundings in the first vertical grid point (critical region for
radiosoundings) we calculated (Table \ref{aws}) the median values of
wind speed and absolute temperature in the three central winter months
obtained from the Automatic Weather Station
(AWS)\footnote{http://amrc.ssec.wisc.edu/realtime.html}. { \bf (An automatic
weather station is an automated version of the traditional weather
station, to enable measurements from remote areas)} and we compared
these outputs with those obtained from radiosoundings in the first
grid point. Data of 2004 are used for South Pole because no AWS data
from 2006 were available. The median wind speed from radiosoundings in
first vertical grid point is 2.8 ms$^{-1}$ and 5.7 ms$^{-1}$
respectively at Dome C and South Pole. These values match in excellent
way with AWS measurements: 3 ms$^{-1}$ and 5.7 ms$^{-1}$ respectively
at Dome C and South Pole in the same periods. We conclude therefore
that the measurement of the first grid point from the radiosoundings
are reliable because perfectly in agreement with AWS measurements.  We
note that a similar median wind speed of 2.6 ms$^{-1}$ has been
calculated at Dome C on a climatologic scale (1984-2003) by \cite{ar}
in winter. \cite{hb}
 calculated at the South
Pole a median wind speed of $\sim$ 5 ms$^{-1}$ in winter on a
climatologic scale (1994-2003).  The median value calculated in our
paper on a time scale of 1 year have therefore almost no climatologic
trends. If we look at the absolute tempearture in Table \ref{aws} we
conclude that the radiosoundings are $\sim$ 1 K colder than the AWS at
Dome C and $\sim$ 2 K at South Pole.

Going back to Fig.\ref{motd} and Fig.\ref{motdd}, this conclusion
supports and confirms the thesis that we are in front of an
overestimate produced by the ECMWF analysis and not to an artifact in
the measurements (radiosoundings). However, the overestimate is
slightly smaller than what predicted for the temperature. This means
an overestimate of 4-5 K instead of 6 K.

Fig.\ref{motd} and Fig.\ref{motdd} therefore indicate that the ECMWF
analyses accurately describe the state of the free atmosphere above
Dome C and the South Pole. In the boundary layer, the ECMWF data shows a
tendency in overestimating the wind speed and the absolute
temperature, particularly in winter. Results we obtained indicated
that ECMWF analyses should be treated with more caution.

These conclusions also confirmed the thesis we had expressed in the
Introduction concerning the \cite{sa} 
paper who claimed
for a good agreement between radiosoundings and ECMWF analyses even in
the surface.

In spite of the fact that they used ERA-reanalyses (a product having a
lower resolution than the MARS catalog used in our study) the
agreement between radiosoundings and analyses in their data matches
well with our findings and predict a good agreement between
radiosoundings and ECMWF analyses in summer. Our analysis, extended to
winter, reveals that, in this season, the agreement is far from being
good and the sharp changes in wind speed and temperature closest to
the surface measured by the radiosoundings are not well reconstructed
by the ECMWF analyses data. 

To provide the most comprehensive and compact comparison of ECMWF
analyses and radiosoundings above Dome C and the South Pole in the
proximity of the surface we prefer to focus now on the two key
parameters defining the thermodynamic stability, i.e. the gradients of
the potential temperature and of the wind speed.  Indeed, only a study
of the simultaneous systematic effects on both quantities can tell us
if we can use ECMWF analyses to quantify the thermodynamic stability
in the surface layer.

Fig. \ref{rs1} shows the median gradient of the potential temperature
(left) and the median of the square of the gradient of the wind speed
(right) in the first 150 m with a vertical resolution of 100 m (a and
b) and of 10 m (c and d). As we can expect, the radiosoundings show a
sharper gradient than the analyses near the surface. We can observe
that the ECMWF analyses are able to identify that the gradients above
Dome C are larger than above South Pole. Unfortunately, a precise
quantification is missing and, even in the case of the best vertical
resolution (cases c and d), the offset produced by analyses with
respect to radiosoundings on the two parameters ($\partial
\theta/\partial z$ and $(\partial v/\partial z)^2$) is not comparable
above the two sites (Dome C and the South Pole). This implies that the
ECMWF analyses do not smooth out the potential temperature and wind
speed gradients in a similar way above the two sites.

Knowing that the Richardson number depends on the ratio of $\partial
\theta/\partial z$ and $(\partial v/\partial z)^2$ we conclude that it
is pretty risky to draw any conclusion on a comparative analyses of
the Richardson number in the surface layer between different sites
calculated with the ECMWF analyses. As a consequence we can state that
we can retrieve a ranking of the three sites with respect to the
thermal and the dynamic gradient in an independent way but we cannot
retrieve a ranking of the three sites with respect to the Richardson
number in the surface layer using ECMWF analyses. We have therefore to
limit us to a comparative analysis of the $\partial \theta/\partial z$
and $(\partial v/\partial z)^2$.  In section 5 we will perform the
Richardson number comparison in the free atmosphere where we showed
the ECMWF analyses are reliable.

For a reliable study of the Richardson number in the surface layer we
need, at present, radiosoundings. A forthcoming paper on this topic is
in preparation.

\subsection{$\partial \theta/\partial z$ and $(\partial v/\partial z)^2$ at the South Pole, Dome C and Dome A}

As a consequence of the conclusions obtained in the previous
section, we present here the 'thermal' and 'dynamic' properties of the
surface layer in an independent way.

The change of the potential temperature with height indicates the
thermal stability of the atmosphere. A positive gradient is defined as
stable conditions, the vertical displacement of air is suppressed and
thus is the production of dynamic turbulence. 

\begin{figure*}
\vbox to160mm{ 
\includegraphics[angle=-90,
    width=17.5cm]{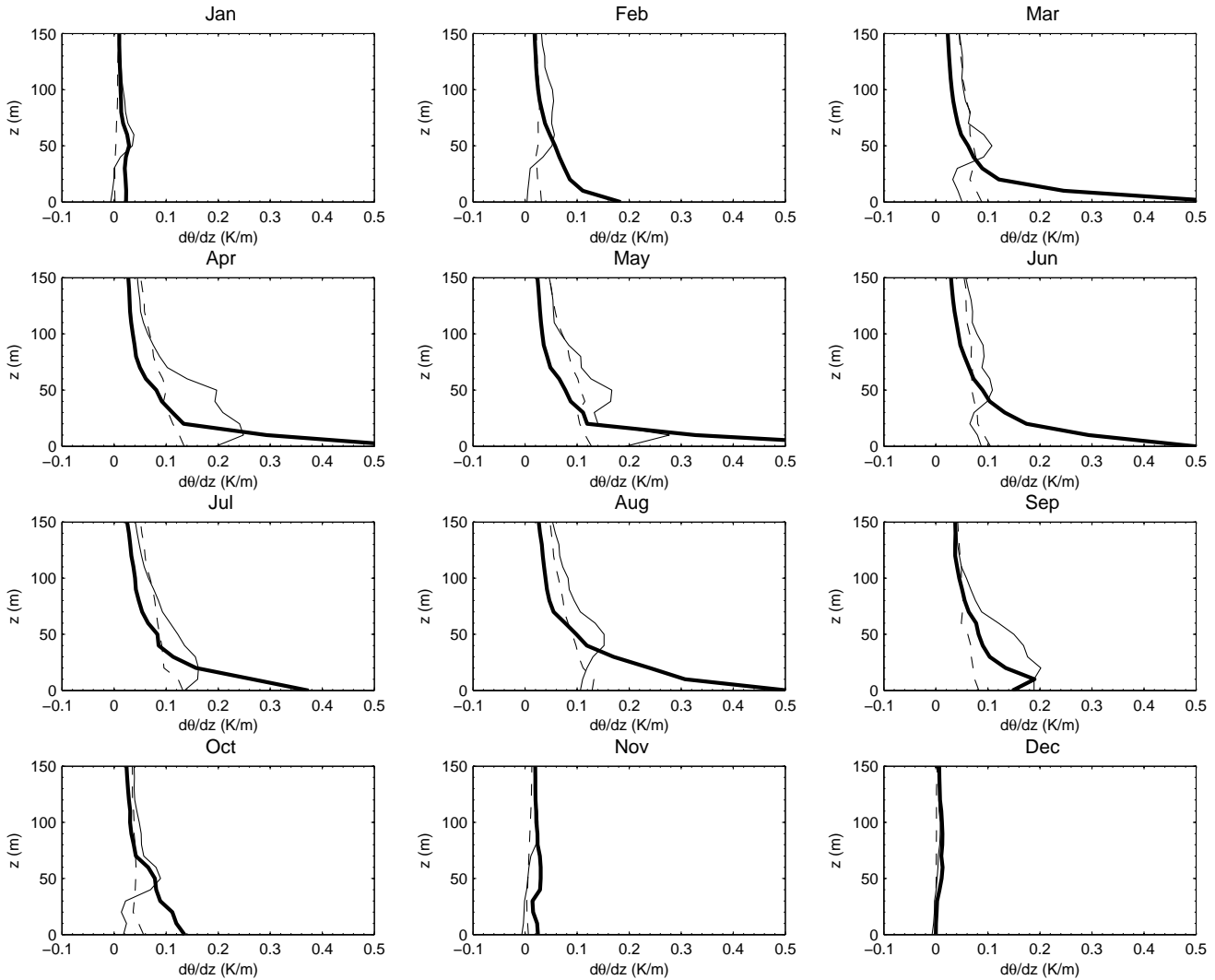}
\caption{The monthly median of the gradient of the potential
  temperature for 2005, Dome A (thick lines), Dome C (thin
  lines) and the South Pole (dashed lines).\label{dth}}
}
\end{figure*}

The absence of sunlight in the antarctic night and the consequent
radiative cooling of the snow surface creates a strong temperature
inversion close to the ground. The monthly median of the gradient of
the potential temperature in the first 150 m for the three sites,
calculated with the ECMWF analyses, is shown in Fig. \ref{dth}. From
February to October all the gradients are positive and indicate a
thermally stable stratified atmosphere near the surface. The inversion
above Dome A (thick lines) is particularly intense when compared to 
Dome C (thin lines) and to the South Pole (dash-dotted
lines) during these months. The value of the gradient at the lowest
level is significantly larger for Dome A almost all the year. From
March to August there is a very sharp change in the slope of the
gradient of Dome A at around 20 m above the surface.

During the summer months all three sites have a neutral stratification
near the surface, i. e. $\partial \theta/\partial z\approx
0$. Vertical motion of the air is not suppressed but neither is it
encouraged. A small perturbation can trigger dynamic turbulence.

\begin{figure*}
\vbox to160mm{
\includegraphics[angle=-90,
    width=175mm]{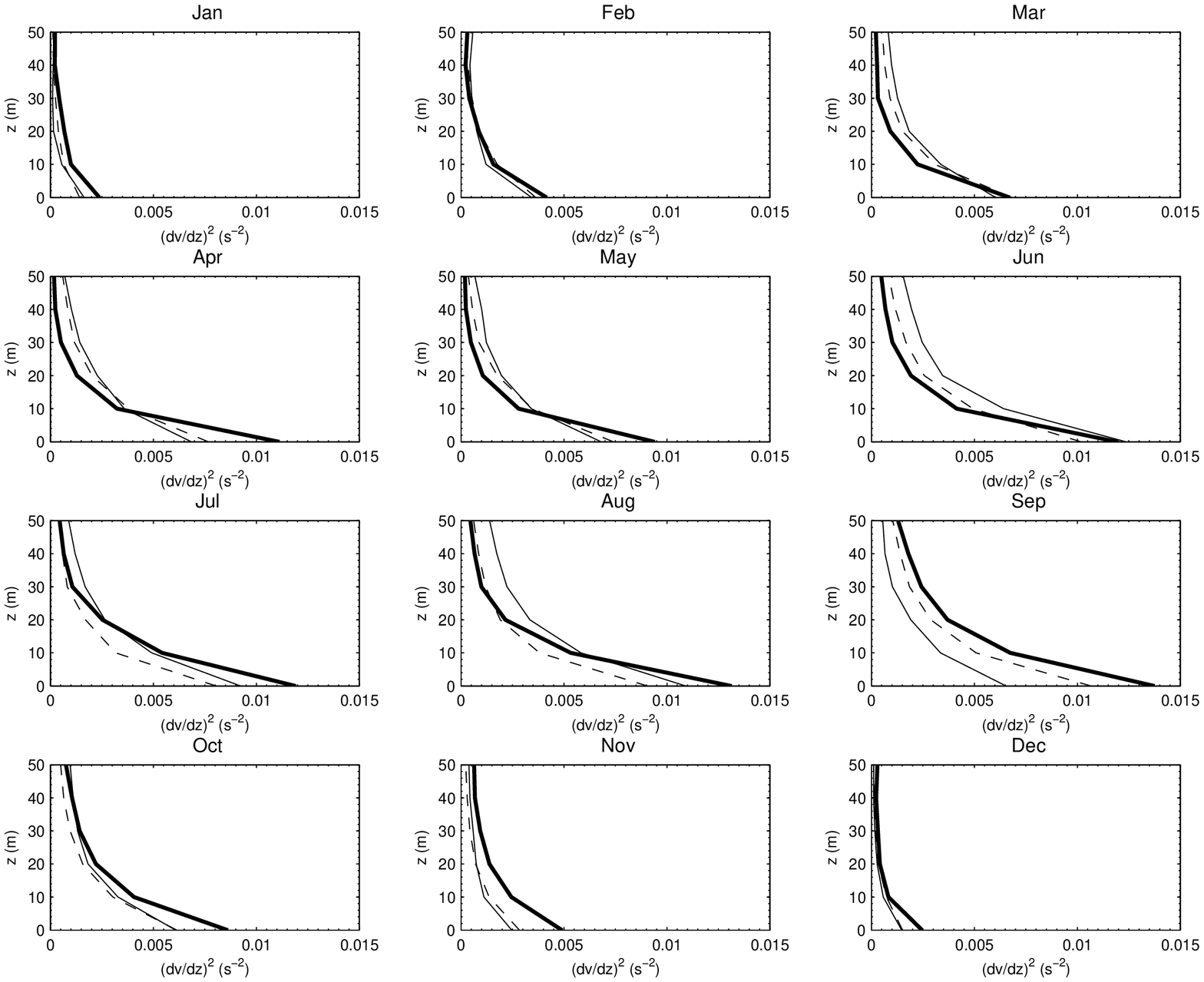} 
\caption{The monthly median of the square of the gradient of the wind
speed in the first 50 m for the year 2005. Dome A (thick lines), Dome
C (thin lines) and the South Pole (dashed lines).\label{gws}}
}
\end{figure*}

The median of the gradient of the square of the wind speed in the
first 50 m above the ground is shown in Fig. \ref{gws}. The
gradient of the wind speed at the lowest level is largest at Dome A
for every month (except in June when it is slightly larger above Dome C).

\section{Radiosoundings: the surface wind speed}
\label{rad}
\begin{figure*}
\vbox to140mm{\includegraphics[angle=-90, width=160mm]{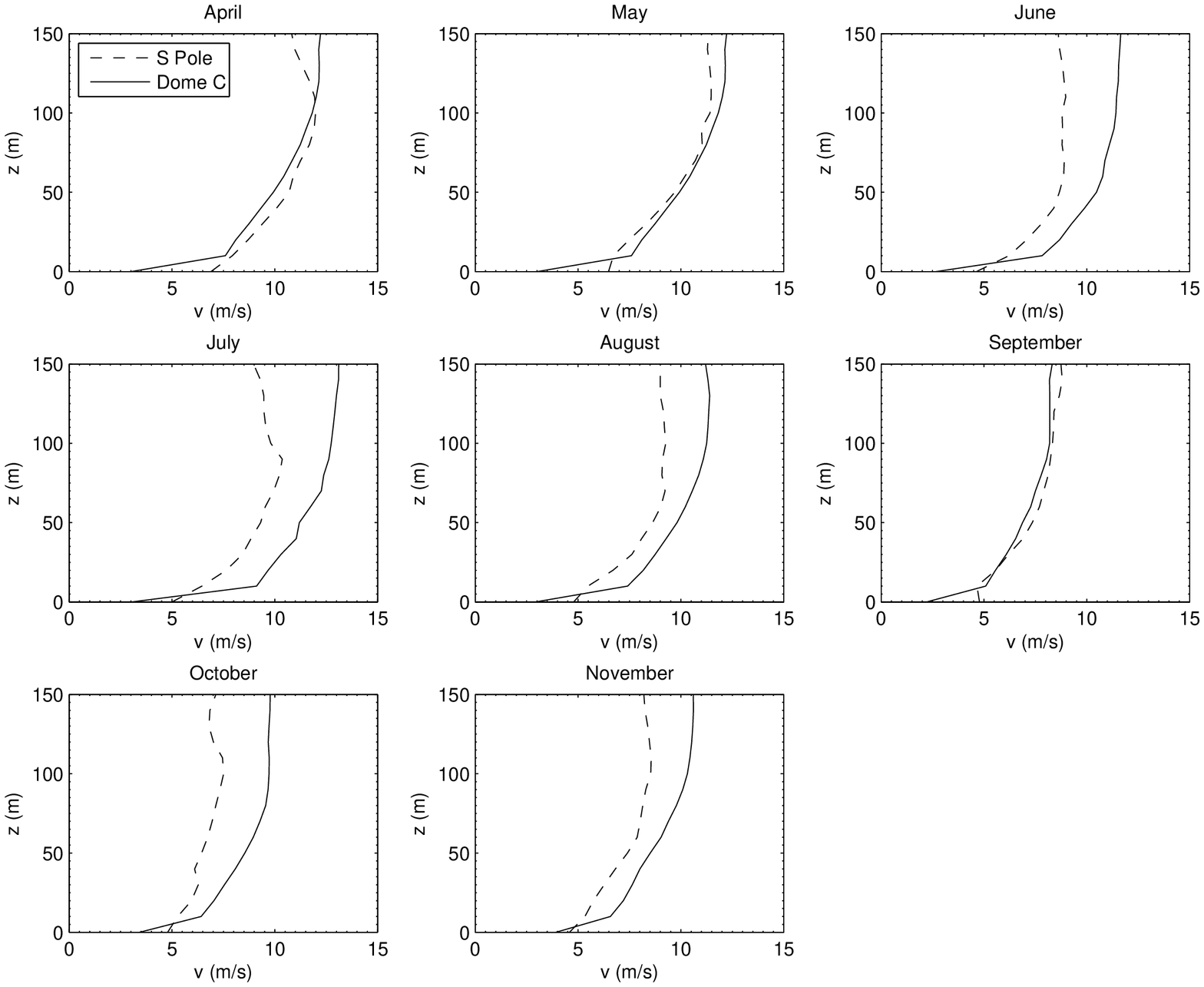}
\caption{The wind speed near the ground at Dome C (solid lines) and the
  South Pole (dashed lines), April to November 2006.\label{rs2}}
}
\end{figure*}

Above the summits of the Internal Antarctic Plateau the surface winds
are expected to be weaker than elsewhere on the Plateau. Fig.\ref{rs2}
shows the median wind speed near the surface measured with
radiosoundings at the South Pole (dashed lines) and Dome C (solid
lines) from April to November. We observed that while it is true that
the wind speed at the very lowest level is weaker at the peak (Dome C)
than at the slope (South Pole), there is a sharp increase in the wind
speed above Dome C in the first few tens of meters. At the height of
10/20 m, from May to November i.e. in the winter, the wind speed is
higher above Dome C than above South Pole.  Above this height the wind
speed at Dome C is either higher than or very similar to the one
observed above the South Pole.

In the center of the winter (June, July and August), the wind speed
above Dome C reaches the 8 ms$^{-1}$ at 20 m and 9 ms$^{-1}$ at 30 m.
 The sharp change of the wind speed in the first 10/20 m matches with
our expectations of a large wind speed gardient. This is indeed a
necessary condition to justify the presence of optical turbulence in
the surface \citep{ag} in spite of a very stable thermal conditions
(i.e. a positive gradient of the potential tempearture).

\cite{trin} 
, in a small sample of
 radiosoundings in winter, observed a wind speed of 5 ms$^{-1}$ at 20
 m and 8 ms$^{-1}$ only at 70 m. Our results, obtained with a complete
 statistical sample in winter, tells us that the Trinquet et
 al. estimate is too optimistic and we should expect a larger wind
 speed at low heights.

We conclude that in the first 10/20 m the mechanical vibrations, that
might be derived from the impact of the atmospheric flow, flowing at
8-9 ms$^{-1}$ on a telescope structure, are probably more critical
above Dome C than above the South Pole and should be carefully taken
into account in the design of astronomical facilities.

\section{ECMWF analyses: the weakest wind speed in the troposphere and
  the lower stratosphere}

\begin{figure*}
\vbox to160mm{ 
\includegraphics[angle=-90, width=175mm]{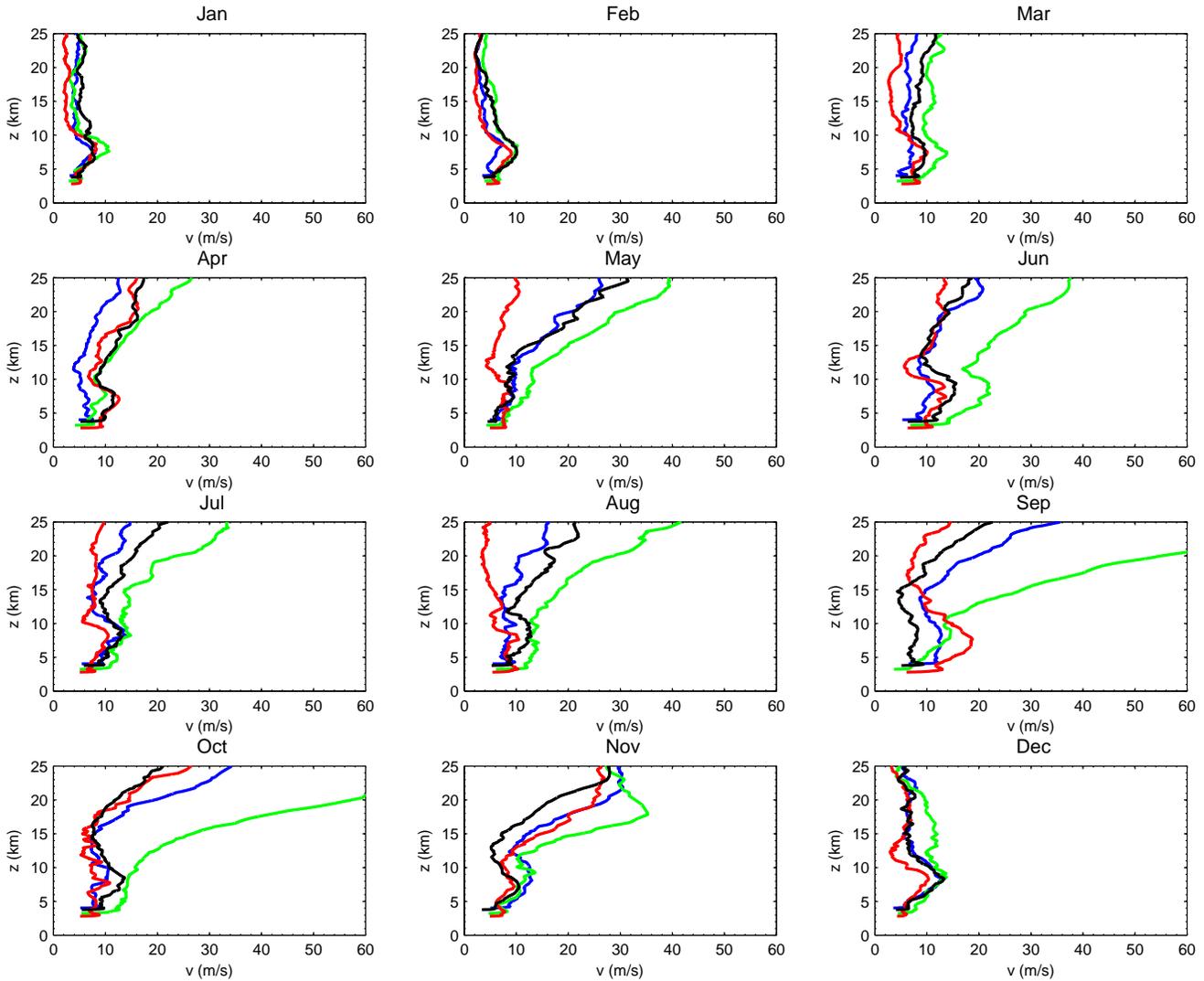} 
\caption{The monthly median wind speed for 2005. Dome
  A (blue line), Dome C (green line), South Pole
  (red line) and Dome F (black line). \label{ws_a}}
}
\end{figure*}

Fig. \ref{ws_a} shows the monthly median of the wind speed from ground
level up to 25 km a.s.l.  for Dome A (blue lines), Dome C (green
lines) and the South Pole (red lines). During summer the wind speed
is quite weak in the upper part of the atmosphere for all three
sites. A maximum is observed at roughly 8 km a.s.l. From December to
April the median wind speed is never larger than 15 m s$^{-1}$ in the
whole 20 km and above all the sites.  As the winter approaches the
wind speed in the upper atmosphere increases. From April to November
the wind speed increases monotonically above 10 km a.s.l. and the
highest wind speed is more likely to be located in the highest part of
the atmosphere. However, the rate at which the wind speed grows up is
not the same above the three sites.  The smallest increase rate is
observed at the South Pole and the maximum rate is found above Dome
C. Differences are far from being negligible. In this slab the median wind
speed is, at Dome C, almost twice that of Dome A or the South Pole.

\cite{gm} found in their investigation a similar wind speed trend
above Dome C from ECMWF analyses for the years 2003 and 2004. The wind
speed trend above Dome C is therefore confirmed in different
years. However, as announced in the Introduction, the goal in our
paper is the comparison of Dome C with other sites to evaluate which
are the best for astronomical applications.

The different wind speed gradient rate observed in Fig. \ref{ws_a}
above different sites is highly probable related to the synoptic
circulation of the Antarctic regions.  Indeed, the jet stream, that
characterizes the vertical wind speed profile of mid-latitude sites at
the tropopause level, is absent here. On the other hand, the Polar
Vortex creates strong high-altitude winds surrounding the Polar High
in winter \cite{sch}. Wind speed increases monotonically above 10 km.
The result observed in Fig.\ref{ws_a} can be explained with the fact
that the South Pole is located near the centre of the polar vortex and
consequently the influence of the polar vortex is weak at this
site. Dome A and Dome C are situated further from the centre of the
continent, and thus from the centre of the polar vortex. It is to be
expected that the wind speed is larger above these sites. The farther
from the polar high (centre of the Polar Vortex) is the site, the
greater is the wind speed strength above 10 km.

\begin{figure}
\vbox to160mm{ 
\includegraphics[width=7cm]{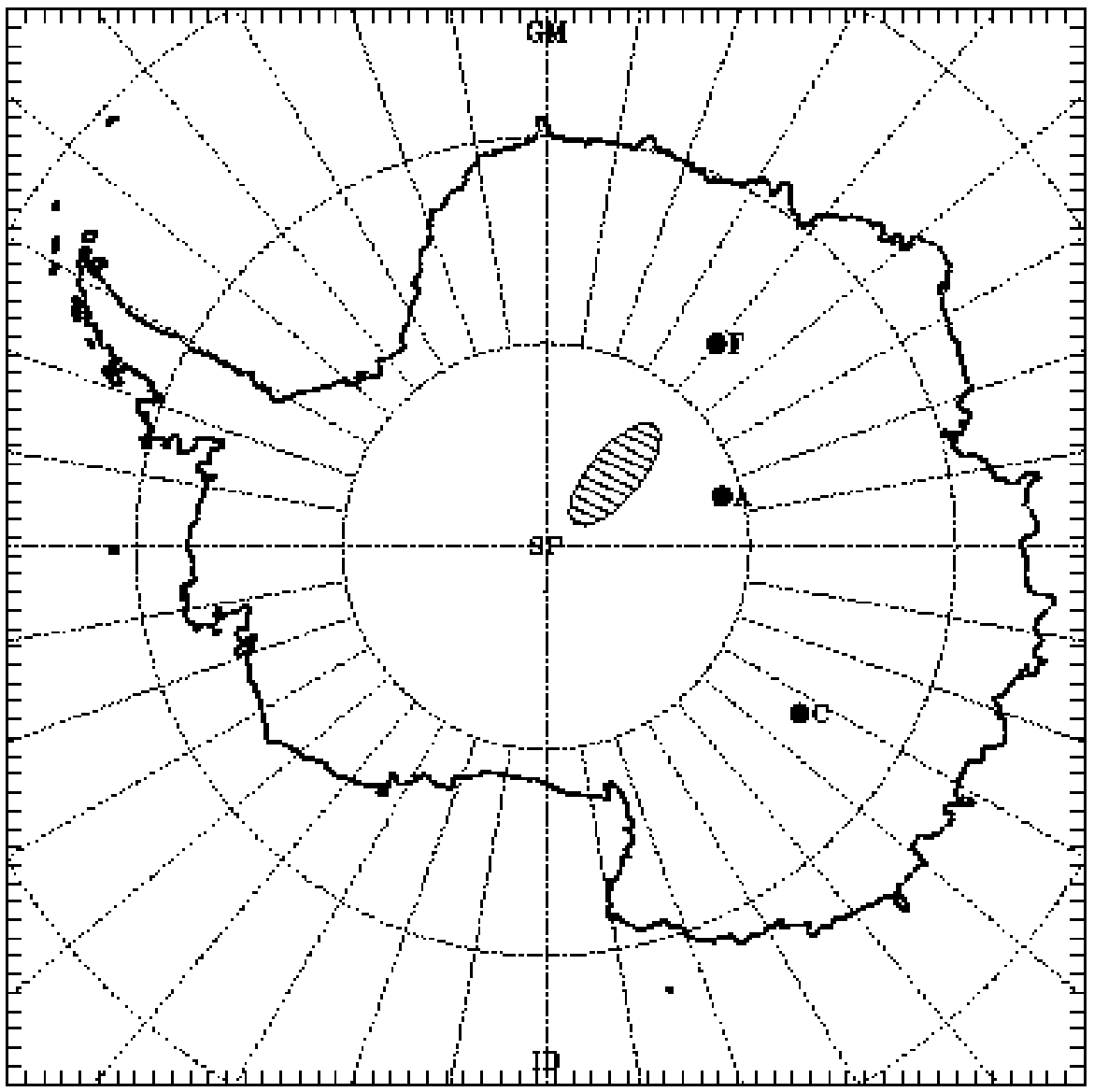} 
\caption{Antarctica map. The sites of South Pole, Dome A, Dome C and
Dome F are labeled with a black point.  The dashed region indicates
the 'position space' of the polar high at different heigths as
retrieved from the Fig. \ref{ws_a}.\label{map}} }
\end{figure}

According to this explanation the wind speed strength in the upper
atmosphere should be related to the distance of the site from the
centre of the polar vortex. If we could know the exact position of the
polar high we would have a perfect tool to identify, {\it a priori},
the site with the weakest wind speed above 10 km in winter.
Unfortunately the Polar Vortex is not a perfect cone and the centre of
the vortex is not located at the same coordinates at all heights.  To
verify the sensitivity to this effect we included in our sample a
further site (Dome F, ($77.31$ S, $39.7$ E), $h$$=$$3810$ m) and we
added the median wind speed in the same Fig.\ref{ws_a}. In this
picture it is evident that South Pole and Dome C have, respectively,
the weakest and the the greatest wind speed above 10 km. The wind
speed above Dome A and Dome F is mostly comparable and it is more
difficult to state which site have the weakest wind speed. Dome A is
slightly better if we consider a statistical analysis of the whole
year but the difference is not so important. This indicates that the
polar high fluctuates probably in a region located between the South Pole,
Dome A and Dome F as indicates the dashed region in Fig.\ref{map}.

\section{The Richardson number for the South Pole, Dome C and Dome A}

The Richardson number is an indicator of the stability of the
atmosphere:

\begin{equation} \label{eRi}
Ri=\frac{g}{\theta}\frac{\partial \theta/\partial z}{(\partial
  v/\partial z)^2}
\end{equation}

where g is the gravitational acceleration (9.8 m s$^{-2}$), $\theta$ is the
potential temperature and v is the wind speed. The atmosphere is
considered to have a stable stratification when the Richardson number
is larger than a critical value, typically 0.25. If the Richardson
number is less than the critical value the stratification is
classified as unstable. The smaller the Richardson number is the
higher is the probability of triggering of turbulence. 

\begin{figure*}
\vbox to160mm{\includegraphics[angle=-90,
    width=175mm]{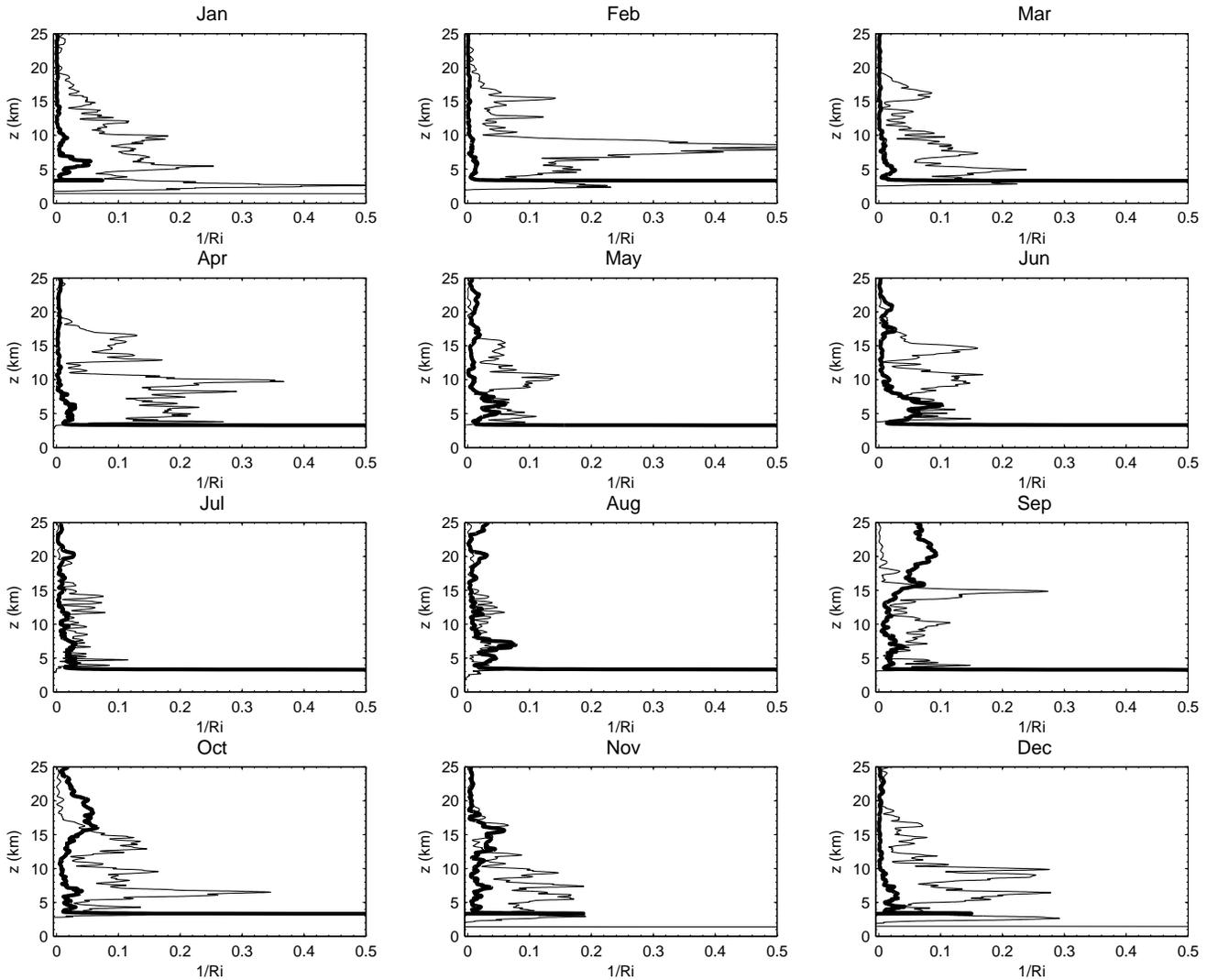}
\caption{The monthly median for 2005 of the inverse of the Richardson number
  (1/Ri) for Dome C (thick solid lines) and Mt. Graham (thin dashed lines).\label{iRi1}}
}
\end{figure*}

The comparison of the Richardson number above different sites allows
us to make a relative estimate of which site is characterized by a
higher or lower probability to trigger turbulence. In
\cite{gm}-Fig. 14 the median of the inverse of the Richardson number
for Dome C are shown in different slabs and periods of the year. From
such results it is possible to retrieve a comparative analysis that
necessarily is qualitative. In other words we can say if a region
shows a higher or smaller probability to trigger turbulence but we do
not have a reference to quantify these differences nor can we conclude
whether these differences are negligible or not. In order to provide insights 
on this question we calculated (Fig. \ref{iRi1}) the median of
the inverse of the Richardson number (1/Ri) from Dome C (thick solid
lines) and from Mt. Graham (thin dashed lines), taken as an example of
a typical mid-latitude site, for each month during the whole year
2005. The inverse of the Richardson number is shown instead of the
Richardson number itself because the inverse shows a better
dynamic. The smaller 1/Ri is, the more stable is the atmosphere. The
1/Ri is smaller above Dome C than above Mt Graham almost everywhere. 
 This result is coherent with the optical turbulence measurements 
observed above the two sites (\cite*{eg}, \cite{ag}) and for this reason it
definitely confirms the method proposed
by \cite{gm} to study 1/Ri as a qualitative relative indicator to rank
sites with respect to the probability to trigger turbulence. During
September and October, when the median wind speed is remarkably strong
at high altitudes at Dome C (see Fig. \ref{ws_a}), 1/Ri is actually
larger than for the mid-latitude site. This confirms, from a
qualitative and quantitative point of view, that the high part of the
atmosphere in September and October is a region to be monitored
carefully because the probability to trigger instabilities is
comparable and even larger than above mid-latitude sites.  A strong
increase in the wind speed is most certainly the cause of the high
1/Ri over Dome C at such altitudes.

\begin{figure*}
\vbox to160mm{\includegraphics[angle=-90,
    width=175mm]{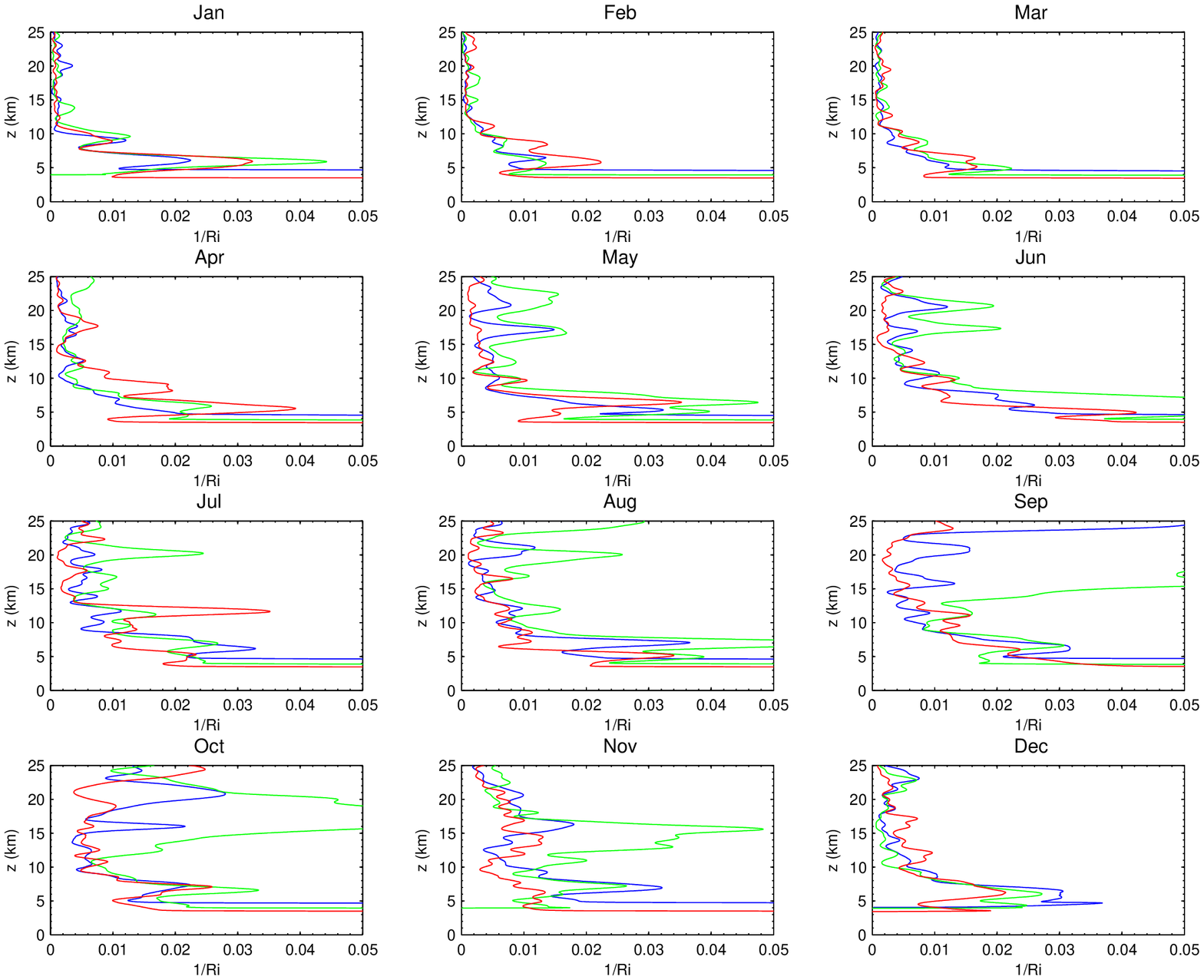} 
\caption{The monthly median for 2005 of the inverse of the Richardson number
  (1/Ri) for Dome A (blue lines), Dome C (green lines) and the South Pole
  (red lines).\label{iRi2}}
}
\end{figure*}

The vertical median profiles of 1/Ri for the three Antarctic sites
(Dome A, Dome C and the South Pole) are shown in Fig. \ref{iRi2}. In
local summer the profiles from the different sites are similar to each
other. The 1/Ri has a maximum at ground level and a smaller peak
somewhere slightly above 6 km a.s.l. Above 10 km a.s.l. the atmosphere
is very stable for all the three sites. From April/May the instability
above 10 km increases. At the end of the winter (September and
October) the instability in the upper part of the atmosphere is even
larger than the maximum value observed near 6 km for Domes A and C in
summer. The instability of Dome C is more pronounced than that of Dome
A. Above the South Pole 1/Ri shows the best conditions (i.e. the most
stable) over the whole year.

\section{Conclusions}
In this paper we provide a first comparison of the atmospheric
characteristics above three different sites on the Internal Antarctic
Plateau: Dome C, Dome A and the South Pole. More precisely we try to
answer the specific questions defined in the introduction.

(1) The comparison of the ECMWF analyses with the radiosoundings shows
that the analyses can accurately describe the atmosphere above the
Internal Antarctic Plateau in the whole range from 10 m to 20 km above
the surface. During no season does the median difference of the wind
speed exceed 1 m s$^{-1}$ above the first 10 m. The median difference
of the absolute temperature is within 2 K in the same vertical
slab. In the surface layer the wind speed discrepancy between analyses
and radiosoundings is slightly larger (2-3 m s$^{-1}$) while ECMWF
analyses show a tendency to overestimate the absolute temperature
measured by radiosoundings in the lowest level in winter ($\Delta$T
$\sim$4-5 K). 
A statistic analysis reveals that most of the
radiosoundings explode in winter at about 10-12 km. This does not
allow us to estimate the reliability of the ECMWF analyses at these
high altitudes.

(2) We proved that the ECMWF analyses do not produce accurate
estimates in the surface layers confirming what was only assumed by
\cite{gm}.  This result represents an answer to our original question
expressed in the Introduction. The ECMWF intrinsically have
limitations for the characterization of the atmospheric flow in this
vertical slab. This justify the employment of meso-scale models but,
at the same time, also tells us that it will be fundamental to prove
that meso-scale models can do better than General Circualtion Models
(GCM) in this vertical slab.

(3) We could conclude that above all the three sites the potential
temperature in the surface layer is extremely stable even if the ECMWF
analyses generally underestimate its gradient when compared to
measurements obtained by radiosoundings. Such an effect is
particularly evident in winter. Dome A is by far the site with the
steepest gradient of potential temperature and wind speed if compared
to the South Pole and Dome C.

(4) We proved that the median wind speed in the first meters above the
ground is weaker at Dome C than at the South Pole from April to
November. However, the wind shear in the surface layer at Dome C is
much larger than at the South Pole achieving at 10-20 m a wind speed
of 8-9 m s$^{-1}$ in winter. Such a strong wind shear combined with a
stable stratification of the air in this layer is most likely to be
the cause of the intense optical turbulence that has been measured in
the first tens of meters at Dome C \citep{ag}. Such a strong wind
speed at this height might be a source of vibrations produced by the
impact of the atmospheric flow on telescope structures and should
therefore be taken into account in the conception and design of
astronomic facilities.

(5) Median monthly values of the inverse of the Richardson number
(1/Ri) indicate that the probability to trigger instabilities is
larger above a mid-latitude site (for which we have a reliable
characterization of the optical turbulence) than above any of the
three sites on the Internal Antarctic Plateau. Above all the three
Antarctic sites 1/Ri is visibly smaller than that measured above
Mt.Graham (selected as representative of mid-latitude sites). This is
the first time that such a conclusion has been achieved and definitely
prove that the method presented in \cite{gm} 
is reliable.

(6) Moreover, our analysis permitted us also a more sophisticated
discrimination between the quality of the the 1/Ri above the three
sites.  Dome A and the South Pole show, indeed, more stable conditions
than Dome C above the first 100 m. This is probably due to the polar
vortex which, producing an increase of the wind speed in the upper
atmosphere, also increases the probability to trigger thermodynamic
instabilities.

(7) We showed that it is risky to retrieve estimates of the Richardson
number in the surface layer (Fig.\ref{rs1}) because we did not find an
equivalent smoothing effect of the ECMWF analyses for the gradient of
the potential temperature nor for the gradient of the wind speed above
different sites.

(8) In the free atmosphere, above the first 10 km, the polar vortex
induces a monotonic increase of the wind speed in winter that is
 proportional to the distance of the site to the polar
high. Dome C therefore shows the largest wind speed above 10 km in
winter. At Dome C the wind speed at 15 km can easily be almost twice
that of Dome A and even thrice the wind speed at the South Pole in
winter. The wind speed above the South Pole is the weakest among the
three sites on the whole 20 km in all seasons. This conclusion put therefore
fundamental warnings for astronomical applications.

(9) This study allowed us to draw a first comprehensive picture of the
atmospheric properties above the Internal Antarctic Plateau.  In spite
of the presence of generally good conditions for astronomical
applications, Dome C does not appear to be the best site with respect
to the wind speed, in the free atmosphere as well as in the surface
layer. Both the South Pole and Dome A show a weaker wind speed in the
free atmosphere. Estimates related to the surface layer need to be
taken with precaution. ECMWF analyses cannot be used to draw
definitive conclusions on comparisons of the three sites in this
vertical slab due to their limited reliability in this thin
atmospheric slab (see Section 5) and radiosoundings are available only
for Dome C and the South Pole. Above Dome A the gradient of the
potential temperature is particularly large in the very near surface
layer indicating conditions of extreme thermal stability that might be
associated to a strong value of the optical turbulence in this
vertical range when a thermodynamic instability occurs (possibly even
larger than above Dome C).  However our study showed that, to predict
the thickness of such a layer we should need measurements or
simulations with atmospheric mesoscale model with a higher spatial
resolution near the ground that is able to better resolve the
evolution of the atmospheric flow. This is a part of our forthcoming
activities.

In conclusion, at present, the real solid and unique argument that makes Dome C
preferable to the South Pole for astronomical applications is the
extreme thinness of the optical turbulence surface layer. We expect at
Dome A comparable or even larger optical turbulence values with
respect to Dome C in the surface layer. We cannot conclude if the
surface layer at Dome A is thinner than that observed above Dome
C. However our study clearly indicates that Dome C is not the best
site on the Internal Antarctic Plateau with respect to the wind speed
in the free atmosphere as well as in the surface layer nor is it the
site with the most stable conditions in the free atmosphere. Both Dome
A and the South Pole show more stable conditions in the free
atmosphere.

\section*{Acknowledgments}
This study has been carried out using radiosoundings from the AMRC
(Antarctic Meteorological Research Center, University of Wisconsin,
Madison $ftp://amrc.ssec.wisc.edu/pub/southpole/radiosonde$) and from
the Progetto di Ricerca 'Osservatorio Meteo Climatologico' of the
Programma Nazionale di Ricerche in Antartide (PNRA),
$http://www.climantartide.it$. ECMWF products are extracted from the
Catalog MARS, $http://www.ecmwf.int$.  This study has been funded by
the Marie Curie Excellence Grant (FOROT) - MEXT-CT-2005-023878.


\label{lastpage}

\begin{thebibliography}{}

\bibitem[\protect\citeauthoryear{Aristidi et al.}{2005}]{ar}
Aristidi, E., Agabi, K., Azouit, M., Fossat, E., Vernin, J., Travouillon, T., Lawrence J.S., Meyer, C., Storey, J.W.V., Halter, B., Roth, W.L., Walden, V., 2005, A\&A, 430, 739
\bibitem[\protect\citeauthoryear{Agabi et al.}{2006}]{ag} Agabi, A.,
Aristidi, E., Azouit, M., Fossat E., Martin F., Sadibekova T., Vernin
J., Ziad A., 2006, PASP, 118, 344
\bibitem[\protect\citeauthoryear{Egner, Masciadri \& McKenna}{Egner et
al.}{2007}]{eg} Egner, S., Masciadri, E., McKenna D., 2007, PASP, 119,
669
\bibitem[\protect\citeauthoryear{Fossat}{2005}]{fos}
Fossat, E. 2005, JApA, 26, 349
\bibitem[\protect\citeauthoryear{Geissler \& Masciadri}{2006}]{gm}
Geissler, K., Masciadri, E., 2006, PASP, 118, 1048
\bibitem[\protect\citeauthoryear{Hudson \& Brandt}{2005}]{hb}
Hudson, S.R. \& Brandt, R.E., 2005, Journ. of Clim., 1673
\bibitem[\protect\citeauthoryear{Lawrence et al.}{2004}]{la}
Lawrence, J., Ashley M., Tokovinin A., Travouillon T, 2004, Nature,
431, 278
\bibitem[\protect\citeauthoryear{Marks}{2002}]{ma}
Marks, R.D., 2002, A\&A, 385, 328
\bibitem[\protect\citeauthoryear{Marks et al.}{1999}]{ma2}
Marks, R.D., Vernin, J., Azouit M., Manigault J.F., Clevelin C., 1999,
A\&AS, 134, 161
\bibitem[\protect\citeauthoryear{Marks et al.}{1996}]{maa}
Marks, R.D., Vernin, J., Azouit, M., Briggs, J.W., Burton, M.G., Ashley, M.C.B., Manigualt, J.F., 1996, 118, 385 
\bibitem[\protect\citeauthoryear{Masciadri, Avila\&Sanchez}{Masciadri
    et al.}{2004}]{b17}
 Masciadri, E., Avila, R., Sanchez, L.J., 2004, RMxAA, 40, 3
\bibitem[\protect\citeauthoryear{Masciadri et al.}{2006}]{forot}
Masciadri, E., Lascaux, F., Stoesz, J., Hagelin, S., Geissler, K., 2007, "Large Astronomical Infrastructures at Concordia, prospects and constraints for Antarctic Optical/IR Astronomy", EAS Publ. Series, 25, 57
\bibitem[\protect\citeauthoryear{Sadibekova et al.}{2006}]{sa}
Sadibekova, T., Fossat, E., Genthon, C., Krinne,r G., Aristidi, E., Agabi,
K., Azouit, M., 2006, Antarctic Science, 18, 437
\bibitem[\protect\citeauthoryear{Schwerdtfeger}{1984}]{sch}
Schwerdtfeger, W., 1984, Weather and climate of the Antarctic, Developments in atmospheric science, 15 (Elsiever)
\bibitem[\protect\citeauthoryear{Swain \& Gall\'ee}{2006}]{sg}
Swain, M. \& Gall\'ee, H., 2006, PASP, 118, 1190 
\bibitem[\protect\citeauthoryear{Storey et al.}{2003}]{sto}
Storey, J.W.V., Ashley, M., C., B., Lawrence, J.S., Burton, M.G. 2003, 
Memorie Sai, 2, 13
\bibitem[\protect\citeauthoryear{Travouillon et al.}{2003}]{tr}
Travouillon, T., Ashley, M.C.B., Burton, M.G., Storey, J. W. V.,
Loewenstein, R.F., 2003, A\&A, 400, 1163
\bibitem[\protect\citeauthoryear{Trinquet et al.}{2008}]{trin}
Trinquet, H., Agabi, K., Vernin, J., Azouit, M., Aristidi, E., Fossat, E., 2008, PASP, accepted
\end{thebibliography}
\end{document}